\documentclass[onecolumn,showpacs]{article}
\usepackage{pdflscape}   
\usepackage{lipsum}      
\usepackage{adjustbox}
\usepackage{graphicx}
\usepackage{dcolumn}
\usepackage{bm}
\usepackage{amsmath}
\usepackage{subcaption}
\usepackage{cite}
\usepackage{amssymb}
\usepackage{multirow}
\usepackage{caption}
\usepackage{epstopdf}
\usepackage{hyperref}
\usepackage{wrapfig}
\usepackage{lipsum} 
\usepackage{cite}

\begin{document}
{\setlength{\oddsidemargin}{1.2in}
\setlength{\evensidemargin}{1.2in} } \baselineskip 0.55cm
\begin{center}
{\LARGE {\bf Interaction of polytropic dark energy in cosmological model: Constraints from observational data}}
\end{center}
\date{\today}
\begin{center}
  Elangbam Chingkheinganba Meetei, S. Surendra Singh \\
   Department of Mathematics, National Institute of Technology Manipur,\\ Imphal-795004,India\\
   Email:{ chingelang@gmail.com, ssuren.mu@gmail.com }\\
\end{center}
\begin{center}
 \textbf{Abstract}
 \end{center}
 We investigate an interacting polytropic dark energy (PDE) model characterised by the equation of state $p_{d} = \alpha \rho_{d}^{\,1+\frac{1}{\beta}}$, where the interaction between dark energy and pressureless matter is modelled via a linear coupling term $Q = 3\eta H\rho_{d}$. The background dynamics are formulated by deriving the Hubble parameter in the interacting scenario, and the model parameters are constrained through a Markov Chain Monte Carlo (MCMC) analysis using three joint observational data sets: Hubble77+BAO26, Hubble77+Pantheon$^+$, and Hubble77+BAO26+DESI DR2. The resulting best-fit values of $(H_0, \Omega_{d0}, \eta) $ are 
$(69.22^{+1.27}_{-1.24},\,0.73^{+0.02}_{-0.02},\,-0.22^{+0.10}_{-0.12})$, $(69.23^{+1.27}_{-1.22},\,0.73^{+0.02}_{-0.02},\,-0.34^{+0.15}_{-0.17})$, and 
$(67.77^{+1.26}_{-1.24},\,0.73^{+0.02}_{-0.02},\,\\-0.02^{+0.10}_{-0.11})$ respectively for the respective data combinations. Our results indicate a positive energy density and negative pressure over the full redshift range, with the evolution of the equation-of-state parameter and state finder parameters placing the model firmly within the Quintessence regime. The study of the deceleration parameter also reveals a shift from a decelerating to an accelerating cosmic expansion. The estimated present age of the Universe is $14\,\mathrm{Gyr}$, consistent with recent observational data. Furthermore, the sign of $Q$ implies a current energy transfer from dark energy to matter. These findings support the interacting PDE framework as a viable candidate for explaining late-time cosmic acceleration and related large-scale dynamics.
\\

 \section{Introduction}\label{Intro}

\hspace{0.5cm}The question of how the Universe originated and evolved remains a central theme in modern cosmology. Current observational and theoretical developments indicate that the Universe emerged approx. $13.8$ Gyr ago from a hot, dense state, as described by the Big Bang paradigm~\cite{ref1}. A brief period of accelerated expansion - inflation - likely occurred within the first fraction of a second, erasing large-scale anisotropies and seeding the primordial density perturbations that later gave rise to cosmic structure. Following recombination at $z \sim 1100$, the cosmic microwave background (CMB) provides a direct snapshot of these initial conditions, while gravitational instability drove the growth of matter over densities into the large-scale structure observed today. Over cosmic time, galaxies, stars, and planets are formed through hierarchical assembly. Observations of Type~Ia supernovae, galaxy clustering, and baryon acoustic oscillations have revealed that the cosmic expansion rate is presently accelerating~\cite{ref2,ref3,Riess}. This acceleration is attributed to a dominant dark energy component, contributing $\sim 68\%$ of the current energy budget, with dark matter and baryonic matter comprising $\sim 27\%$ and $\sim 5\%$, respectively. Dark energy’s physical nature remains unknown, while dark matter has been inferred only through its gravitational effects---rotation curves, gravitational lensing, and CMB anisotropies. Together, these components form the basis of the $\Lambda$CDM model, which is highly successful yet faces persistent tensions, notably the discrepancy in local and early-Universe measurements of the Hubble constant~\cite{Planck,ref5,ref54,ref6,ref7}. Such issues have motivated extensions to the standard framework, including modified gravity theories and non-standard matter sectors.

A well-explored extension involves dynamical dark energy models, in which the dark energy density evolves with cosmic time and may exchange energy or momentum with dark matter. Interacting scenarios have been studied in various gravitational frameworks, including $f(R,T^{\phi})$~\cite{Amit1}, $f(T,B)$~\cite{Amit2}, and anisotropic cosmologies~\cite{Shivangi1}, as well as nonlinear polytropic gas models~\cite{Khurshudyan}. These interactions can modify the late-time expansion rate without invoking a finely tuned cosmological constant~\cite{ref23,ref24}, potentially easing the cosmological constant problem. Scalar-field realizations of such models allow the coupling strength to depend on the kinetic and potential terms of the field~\cite{ref25}.

Another approach to solve the late time cosmic acceleration is the modification of general relativity. Modified gravity also able incorporate the presence of dark energy. For instance,~\cite{ref10} examined cosmological models extending beyond the standard model,~\cite{ref11} addressed technical aspects of spherical systems coupled to isotropic matter and horizon issues in higher-curvature gravity, and~\cite{ref12,ref14,ref17} discussed solar system constraints on $f(G)$ gravity, extra forces in $f(R)$ gravity, and unified cosmic histories in $f(R)$ gravity and Lorentz non-invariant models. In the context of modified gravity,~\cite{Amit3} studied the cosmological dynamics and thermodynamic behaviour of $f(Q,C)$ gravity,~\cite{ref20,ref21} briefly examined $f(R,T)$ gravity, and~\cite{ref29} considered $f(Q,T)$ gravity with a $q(z)$ parametrization.  

Among various cosmological models, the polytropic dark energy model has attracted considerable attention due to its mathematical simplicity and its capacity to unify different cosmic epochs within a single framework. In addition, the polytropic dark energy model has been found to accommodate diverse observational constraints while remaining compatible with the standard hot Big Bang framework. Its phenomenological nature allows for its use in interacting dark sector models, scalar field reconstructions, and extensions to modified gravity theories. As a result, the polytropic EoS serves as a versatile and physically motivated framework for exploring alternative descriptions of the Universe’s accelerated expansion.

  In \cite{Karami} authors investigated the polytropic gas model using cosmological observations, while in \cite{Safae} authors explored its implications for late-time cosmic acceleration through an observational approach. In \cite{Salti} authors provided a brief discussion on the variable polytropic gas cosmology. However, interactions between dark energy and dark matter may significantly influence the cosmic expansion rate. Recent researches have increasingly focused on such interactions as a potential means to address the cosmological constant problem~\cite{ref23}. Unlike the standard Cold Dark Matter (CDM) framework, interacting dark energy models allow for energy or momentum exchange between these components~\cite{ref24}, potentially altering the expansion history without invoking a finely tuned cosmological constant. In scalar field models of dark energy, the coupling to dark matter depends on the scalar field’s kinetic and potential energy, both of which critically affect the acceleration rate~\cite{ref25}. This interaction-driven perspective improves the flexibility of cosmological models, offering a pathway to overcome theoretical challenges and observational tensions present in non-interacting scenarios. 

To test such models, we employ the polytropic dark energy framework. Here, we analyze the interacting polytropic dark energy model within the context of general relativity in conjunction with observational constraints, aiming to assess its consistency with current data and its potential to address cosmological tensions. Motivated by these above discussions, we adopt in this work a simple linear interaction term between dark matter and dark energy. This allows for a bidirectional transfer of energy, either from dark matter to dark energy or vice versa.

In this work, we investigate the interaction of the polytropic form of dark energy in the framework of general relativity. The paper is organized as follows: section~\ref{poly} provides an overview of the polytropic form of dark energy. Section~\ref{EFEs} presents the Einstein field equations (EFEs) and the corresponding cosmological solutions in an interacting scenario. Section~\ref{observation} discusses the observational constraints on the model parameters. Section~\ref{cosmic parameters} examines various cosmic parameters, including the effective pressure, effective energy density, equation of state parameter, deceleration parameter and density parameter. Section~\ref{sf} presents the state-finder diagnostic analysis. Section~\ref{age} examines the age of the Universe. Section~\ref{behaviour} discusses about the behaviour of the interacting term. Finally, section~\ref{dis} contains the concluding discussion.

\section{Overview of polytropic form of dark energy}\label{poly}

\hspace{0.5cm}The polytropic form of dark energy is a phenomenological cosmological framework in which the equation of state (EoS) is motivated by the well-known polytropic relation from fluid dynamics and astrophysics. In its general form, the polytropic EoS can be written as
\begin{equation}\label{form}
p = \alpha \rho^{1 + \frac{1}{\beta}},
\end{equation}
where $p$ is the pressure, $\rho$ is the energy density, $\alpha$ is the polytropic constant, and $\beta$ is the polytropic index. This EoS generalizes the perfect fluid form $p = \omega \rho$ by introducing a nonlinear dependence of the pressure on the energy density. The parameter $\alpha$ governs both the sign and magnitude of the pressure, thereby determining whether the fluid behaves in a repulsive (accelerating) or attractive (decelerating) manner, while the parameter $\beta$ controls the degree of nonlinearity and allows for richer dynamical behaviour compared to a constant-$\omega$ model.

From a cosmological perspective, the polytropic form has several appealing features. First, for certain positive values of $\beta$, the effective pressure becomes negative at low redshifts, enabling the model to account for the observed late-time accelerated expansion of the Universe. In the limit $\beta \to \infty$, Eq.~\eqref{form} reduces to the constant-$\omega$ dark energy EoS, which for $\omega = -1$ recovers the cosmological constant $\Lambda$ of the $\Lambda$CDM model. For $\alpha = 0$, the model describes a dust-like pressureless matter component ($p=0$), corresponding to the standard cold dark matter case. 

The evolution dictated by Eq.~\eqref{form} naturally interpolates between different cosmic regimes: at early times (high $\rho$), the fluid behaves as pressureless matter, while at late times (low $\rho$) it may generate sufficient negative pressure to drive acceleration. This dynamical transition can, depending on the values of $\alpha$ and $\beta$, yield an effective EoS parameter $\omega_{\mathrm{eff}} = p/\rho$ that is: Quintessence-like ($-1 < \omega_{\mathrm{eff}} < -1/3$), producing accelerated expansion without crossing the phantom divide, phantom-like ($\omega_{\mathrm{eff}} < -1$), potentially leading to a future singularity such as the Big Rip, or Cosmological-constant-like ($\omega_{\mathrm{eff}} \approx -1$), mimicking $\Lambda$CDM.

Such flexibility allows the polytropic EoS to encompass a wide range of cosmological scenarios within a single unified formalism. An important advantage of this approach is that it offers the possibility of unifying dark matter and dark energy into a single effective cosmic fluid. In such a unification scheme, the early-time dust-like behaviour accounts for the clustering properties of dark matter, while the late-time negative pressure behaviour explains cosmic acceleration, all without introducing additional exotic scalar fields or modifications to gravity. Moreover, the ability of the parameter $\beta$ to regulate the transition between these regimes provides a mechanism to alleviate both the fine-tuning problem (by avoiding the need to set the vacuum energy density to an extremely small value) and the cosmic coincidence problem (by naturally linking the onset of acceleration to the matter-dominated era).

 \section{Einstein field equations (EFEs) and cosmological solutions in an interacting scenario}\label{EFEs}
 
 \hspace{0.5cm}In the current section, we are working on the FLRW metric of a flat, homogeneous and isotropic Universe filled with dark matter and dark energy. Its corresponding metric in cartesian coordinates is given by the following equation
 \begin{equation}\label{metric}
   ds^{2}=-dt^{2}+a(t)\sum_{i=1}^{3}dx_{i}^{2}
 \end{equation}
 where $a(t)$ is the scale factor at cosmic time $t$. The corresponding EFEs are given by
 \begin{equation}\label{fieldEs}
 \begin{split}
    3H^2 & = \rho_{m}+\rho_{d} \\
     2\dot{H} & =-( \rho_{m}+p_{m}+\rho_{d}+p_{d}),
 \end{split}
 \end{equation} 
 where $H=\frac{\dot{a}}{a}$ is the Hubble parameter, $\rho_{m}$ and $\rho_{d}$ represent the energy density of the dark matter and dark energy respectively. While $p_{m}$ and $p_{d}$ denote the pressure of the dark matter and dark energy respectively. Here, we are considering a pressureless dark matter so that $p_{m}=0$, implying $\omega_{m}=0$. While for that of the dark energy, we are considering a polytropic type of dark energy, by using equation (\ref{form}) its corresponding EoS parameter is given below as 
 \begin{equation}\label{formEOS}
   \omega_{d}=\alpha \rho_{d}^{\frac{1}{\beta}}.
 \end{equation}
 
 The above EFEs in eqn. (3) can be written as 
 \begin{equation}\label{fieldEseff}
 \begin{split}
    3H^2 & = \rho_{eff} \\
     2\dot{H} & =-(P_{eff}+\rho_{eff}),
 \end{split}
 \end{equation} 
 where $\rho_{eff}=\rho_{m}+\rho_{d}$ and $P_{eff}=p_{d}$. The basic conservation equations are given by
 \begin{equation}\label{conservation}
 \begin{split}
     & \dot{\rho_{m}}+3H\rho_{m}=0 \\
      & \dot{\rho_{d}}+3H\rho_{d}(1+\omega_{d})=0\\
      & \dot{\rho}_{eff}+3H\rho_{eff}(1+\omega_{eff})=0     
 \end{split}
 \end{equation}
 
 The effective EoS and deceleration parameters are given by
 \begin{equation}\label{effeos}
 \omega_{eff}=\frac{\rho_{eff}}{P_{eff}}
 \end{equation}
 and
 \begin{equation}\label{decc}
 q=-1-\frac{\dot{H}}{H^2}
 \end{equation}
 
 The physical nature of dark energy remains largely uncertain. In the standard non-interacting $\Lambda$CDM framework, dark energy maintains a constant density, whereas the dark matter density evolves as $\rho_m \propto (1+z)^3$. Consequently, the fact that their energy densities are of comparable magnitude at the present epoch appears to be a remarkable coincidence, often referred to as the cosmic coincidence problem. To address this, we investigate a possible interaction between dark matter and dark energy via an interaction term $Q$, which may offer a better agreement with certain large-scale structure 
observations. In this work, we adopt a linear form of the interaction, given by 
\begin{equation}
    Q = 3\eta H \rho_{d},
\end{equation}
where $\eta$ is the dimensionless coupling constant. Thus the modified conservation equations are given by 
 
 \begin{equation}\label{conservation_Eq}
      \dot{\rho_{m}}+3H\rho_{m}=-Q 
      \end{equation}
      and
      \begin{equation}\label{con_eq1}
       \dot{\rho_{d}}+3H\rho_{d}(1+\omega_{d})=Q.
 \end{equation}
 
 If $Q$ is negative i.e. $\eta<0$, energy flows from dark energy to dark matter while if $Q$ is positive i.e. $\eta>0$, energy flows from dark matter to dark energy. Thus, using the relation $a=\frac{1}{1+z}$, equations (\ref{conservation_Eq}) and (\ref{con_eq1}) can be written as
 \begin{equation}\label{eocMatter}
      (1+z)\rho_{m}' -3\rho_{m}  = 3\eta\rho_{d} 
      \end{equation}
 and
      \begin{equation}\label{eocEnergy}
       (1+z)\rho_{d}'-3\rho_{d}(1+\alpha\rho_{d}^{\frac{1}{\beta}})  = -3\eta\rho_{d}.
 \end{equation}
 
 Here $(')$ denotes derivative with respect to $z$. Solving equation (\ref{eocEnergy}), we get 
 \begin{equation}\label{rhoD}
   \rho_{d}=\rho_{d0}\left[\left(1+K\right)(1+z)^{\frac{3(\eta-1)}{\beta}}-K\right]^{-\beta}
 \end{equation}
 where $K=\frac{\alpha}{1-\eta}\rho_{d0}^{\frac{1}{\beta}}$ and $\rho_{d0}$ is the present value of $\rho_{d}$. Using equation (\ref{rhoD}), equation (\ref{eocMatter}) can be written as
 \begin{equation}\label{rhoM}
   \frac{d\rho_{m}}{dz}-\frac{3\rho_{m}}{1+z}=3\eta(1+z)^{-1}\rho_{d0}\left[\left(1+K\right)(1+z)^{\frac{3(\eta-1)}{\beta}}-K\right]^{-\beta}
 \end{equation}
 
 The above equation is a first order linear differential equation and its corresponding integrating factor is $(1+z)^{-3}$. Thus, equation (\ref{rhoD}) can be written as 
 \begin{equation}\label{rhoM1}
   \frac{d}{dz}\left(\frac{\rho_{m}}{(1+z)^3}\right)=\int3\eta(1+z)^{-4}\rho_{d0}\left[\left(1+K\right)(1+z)^{\frac{3(\eta-1)}{\beta}}-K\right]^{-\beta}dz
 \end{equation}
  
  The above equation is difficult  to solve with normal integration. So, letting $\frac{3(\eta-1)}{\beta}=-3, \quad \text i.e. \quad \beta=1-\eta$, the solution of the above differential equation is obtained as
  \begin{equation}\label{rhoM2}
    \rho_{m}=\rho_{m0}(1+z)^{3}\left[\left(1+K\right)(1+z)^{-3}-K\right]^{\eta}.
  \end{equation}
 where $\rho_{m0}$ is the present value of $\rho_m$. Also, under the condition $\beta=1-\eta$, equation (\ref{rhoD}) can be rewritten as
 \begin{equation}\label{rhoD1}
   \rho_{d}=\rho_{d0}\left[\left(1+K\right)(1+z)^{-3}-K\right]^{\eta-1}
 \end{equation}
 
 Density parameter of the dark matter and dark energy are defined as
 \begin{equation}\label{densityparameter}
   \Omega_{m}=\frac{\rho_{m}}{3H^{2}}, \quad  \Omega_{d}=\frac{\rho_{d}}{3H^{2}}. 
 \end{equation}
 
  By equation (\ref{EFEs}), we have
  \begin{equation}\label{1}
    1=\Omega_{m}+\Omega_{d}.
      \end{equation}
  
  Thus, using equations (\ref{rhoM2}) and (\ref{rhoD1}) in equation (\ref{fieldEs}), we obtained the Hubble parameter as
  \begin{equation}\label{Hubble}
    H(z)=H_{0}\left[(1-\Omega_{d0})(1+z)^{3}\left[\left(1+K\right)(1+z)^{-3}-K\right]^{\eta}+\Omega_{d0}\left[\left(1+K\right)(1+z)^{-3}-K\right]^{\eta-1}\right]^{1/2}
  \end{equation}
    where $H_{0}$ and $\Omega_{d0}$ are the present values of $H(z)$ and $\Omega_{d}$ respectively.
  
  \section{Observational constraints}\label{observation}
 \hspace{0.5cm}In the present section, the model parameters $H_{0}$, $\Omega_{d0}$, $\eta$, and $K$ are constrained using the MCMC approach. This method is well suited for efficiently exploring multidimensional parameter spaces and obtaining statistically robust estimates with well-defined confidence intervals. In the subsequent section, these best-fit values will be employed to examine various cosmological parameters and physical properties of the Universe. The observational datasets used for parameter estimation include $77$ Hubble parameter measurements, $26$ BAO measurements, DESI DR2 BAO points, and the Pantheon$^{+}$ compilation of $1701$ Type Ia supernovae. Together, these datasets provide a comprehensive probe of the cosmic expansion history, enabling stringent and reliable constraints on the model.

\begin{center}
\textbf{A. Hubble data}
\end{center}
\hspace{0.5cm}The Hubble parameter data provide direct information about the expansion rate of the Universe and play a crucial role in reconstructing its expansion history, thereby offering a powerful means of testing cosmological models, particularly those extending beyond the standard $\Lambda$CDM framework. The Hubble parameter, $H(z)$, is measured as a function of redshift using various observational techniques, among which the most common are:  

\begin{enumerate}
    \item Differential Age Method (Cosmic Chronometers): This method estimates $H(z)$ by determining the age difference between passively evolving galaxies at different redshifts. It is based on the relation
    \begin{equation}
    H(z) = -\frac{1}{1+z} \frac{dz}{dt},
    \end{equation}
    where stellar population synthesis models are used to estimate the differential age $dt$, and spectroscopic observations provide the redshift variation $dz$.
    
    \item Baryon Acoustic Oscillations (BAO): This technique exploits the imprint of sound waves from the early Universe, which left a characteristic scale in the large-scale distribution of matter. By measuring this BAO scale at different redshifts, one obtains a standard cosmological ruler, enabling determination of $H(z)$ \cite{ref33,ref34}.
\end{enumerate}

In the present analysis, we employ a compilation of $77$ measurements of the Hubble parameter in the redshift range $z \in [0,\, 2.36]$, as presented in \cite{ref30,ref31}. The model parameters $H_0$, $K$, $\eta$, and $\Omega_{d0}$ are estimated using the maximum likelihood approach, which is equivalent to minimizing the chi-square statistic. The chi-square function for the Hubble data is given by
\begin{equation}
\chi^2_H = \sum_{i=1}^{77} \frac{\left[ H_{\mathrm{th}}(p, z_i) - H_{\mathrm{obs}}(z_i) \right]^2}{\left[ \sigma_H(z_i) \right]^2},
\end{equation}
where $H_{\mathrm{th}}(p, z_i)$ denotes the theoretical prediction from the model at redshift $z_i$, $p$ represents the set of free parameters to be fitted, $H_{\mathrm{obs}}(z_i)$ is the observed value, and $\sigma_H(z_i)$ is the corresponding observational uncertainty.

\begin{center}
\textbf{B. BAO dataset}
\end{center}

\hspace{0.5cm}Baryon Acoustic Oscillations (BAO) originate from the residual imprints of sound waves that propagated through the photon–baryon plasma in the early Universe. Prior to recombination, radiation pressure supported oscillations in this tightly coupled fluid. As the Universe expanded and cooled, recombination occurred, photons decoupled from matter, and these oscillations became “frozen” into the large-scale distribution of matter. The physical scale associated with this process, known as the sound horizon, is well constrained from Cosmic Microwave Background (CMB) observations and is typically approximated as $r_d \approx 147\,\mathrm{Mpc}$. This characteristic comoving length serves as a standard ruler for probing cosmological distances and the expansion history.

In this work, we employ a compilation of $26$ BAO measurements, including the most recent DESI DR2 results, derived from galaxy redshift surveys and Lyman-$\alpha$ forest analyses \cite{ref49,ref50,ref51,ref52}. These data constrain the cosmic expansion history by comparing model predictions of various cosmological distance estimators to the BAO scale. The key BAO quantities considered in our analysis are:

\begin{enumerate}
    \item Transverse comoving distance ratio: This quantity relates angular separations on the sky to cosmological distances and is defined as
    \begin{equation}
    \frac{d_M(z)}{r_d} = \frac{D_L(z)}{r_d(1+z)},
    \end{equation}
    where $d_M(z)$ is the transverse comoving distance and $D_L(z)$ is the luminosity distance.

    \item Hubble distance ratio: This parameter characterises the expansion rate of the Universe along the line of sight and is given by
    \begin{equation}
    \frac{d_H(z)}{r_d} = \frac{c}{r_d H(z)},
    \end{equation}
    where $d_H(z)$ is the Hubble distance.

    \item Volume-averaged distance ratio: To obtain an effective isotropic BAO measurement that combines both radial and transverse information, we use the volume-averaged distance ratio,
    \begin{equation}
    \frac{d_V(z)}{r_d} = \left[\frac{z\, d_M^2(z)\, d_H(z)}{r_d^3}\right]^{1/3}.
    \end{equation}
\end{enumerate}

The agreement between theoretical predictions and BAO observations is quantified via the chi-square statistic:
\begin{equation}
\chi^2_{\mathrm{BAO}} = \sum_{k=1}^{N} \left( \frac{Y_k^{\mathrm{th}} - Y_k^{\mathrm{obs}}}{\sigma_{Y_k}} \right)^2,
\end{equation}
where $Y_k^{\mathrm{th}}$ and $Y_k^{\mathrm{obs}}$ denote the theoretical and observed values, respectively, of a BAO quantity at redshift $z_k$, and $\sigma_{Y_k}$ is the corresponding measurement uncertainty. This statistical framework enables robust constraints on cosmological parameters by evaluating the consistency between the predicted and observed large-scale structure. The full list of the $26$ BAO data points employed in this work is provided in \cite{ref31}.

\begin{center}
\textbf{C. Pantheon$^{+}$ datset}
\end{center}
\hspace{0.5cm}The Pantheon$^{+}$ compilation \cite{ref30,ref31} comprises $1701$ spectroscopically confirmed Type Ia supernovae (SNe Ia) spanning the redshift range $0.001 < z < 2.26$. This dataset represents one of the most extensive and homogeneous SN Ia samples currently available, providing a powerful tool for constraining the cosmic expansion history. The high-precision measurements of luminosity distances, and the corresponding comoving distances, enable stringent tests of cosmological models and allow for detailed investigations into the time evolution of cosmic acceleration. The comoving distance, $D_C(z)$, is a fundamental cosmological quantity representing the proper separation between two objects at a fixed cosmic time, corrected for the effect of cosmic expansion. In a spatially flat universe, it is defined as
\begin{equation}
    D_C(z) = c \int_0^z \frac{dz'}{H(z')},
\end{equation}
where $c \approx 3.0 \times 10^8\,\mathrm{m\,s^{-1}}$ is the speed of light, and $H(z)$ is the Hubble parameter at redshift $z$. The comoving distance encodes the cumulative effect of the Universe’s expansion rate and is thus sensitive to the relative contributions of matter, radiation, and dark energy at different epochs. Another key observable is the luminosity distance, $D_L(z)$, which relates the intrinsic luminosity of an astrophysical source to its observed flux. For standard candles such as SNe Ia, $D_L(z)$ is related to the comoving distance by
\begin{equation}
    D_L(z) = D_C(z)\,(1+z).
\end{equation}
This relation reflects the additional $(1+z)$ factor due to the redshifting of photons and the time dilation of their arrival rate. The luminosity distance is commonly expressed through the distance modulus, $\mu(z)$, which quantifies the difference between an object’s apparent and absolute magnitudes:
\begin{equation}
    \mu(z) = 5 \log_{10} \left( \frac{D_L(z)}{\mathrm{Mpc}} \right) + 25.
\end{equation}
Here, $D_L$ is given in megaparsecs (Mpc), and the additive constant $25$ arises from the logarithmic definition of astronomical magnitudes. To compare theoretical models with the Pantheon$^{+}$ observations, we define the chi-square statistic as
\begin{equation}
    \chi^2_{\mathrm{Pantheon+}} = \sum_{i=1}^{1701} \frac{\left[ \mu_{\mathrm{obs}}(z_i) - \mu_{\mathrm{th}}(z_i) \right]^2}{\sigma^2_\mu(z_i)},
\end{equation}
where $\mu_{\mathrm{obs}}(z_i)$ is the observed distance modulus at redshift $z_i$, $\mu_{\mathrm{th}}(z_i)$ is the model-predicted value, and $\sigma_\mu(z_i)$ is the associated observational uncertainty. For the combined cosmological analysis, the total chi-square function is given by
\begin{equation}
    \chi^2_{\mathrm{tot}} = \chi^2_{H(z)} + \chi^2_{\mathrm{BAO}} + \chi^2_{\mathrm{Pantheon+}},
\end{equation}
where $\chi^2_{H(z)}$ and $\chi^2_{\mathrm{BAO}}$ represent the chi-square contributions from Hubble parameter and BAO measurements, respectively. This joint statistical framework enables simultaneous constraints on model parameters using multiple complementary cosmological probes.
  
  \begin{figure}[hbt!]
  \centering
    \includegraphics[scale=0.3]{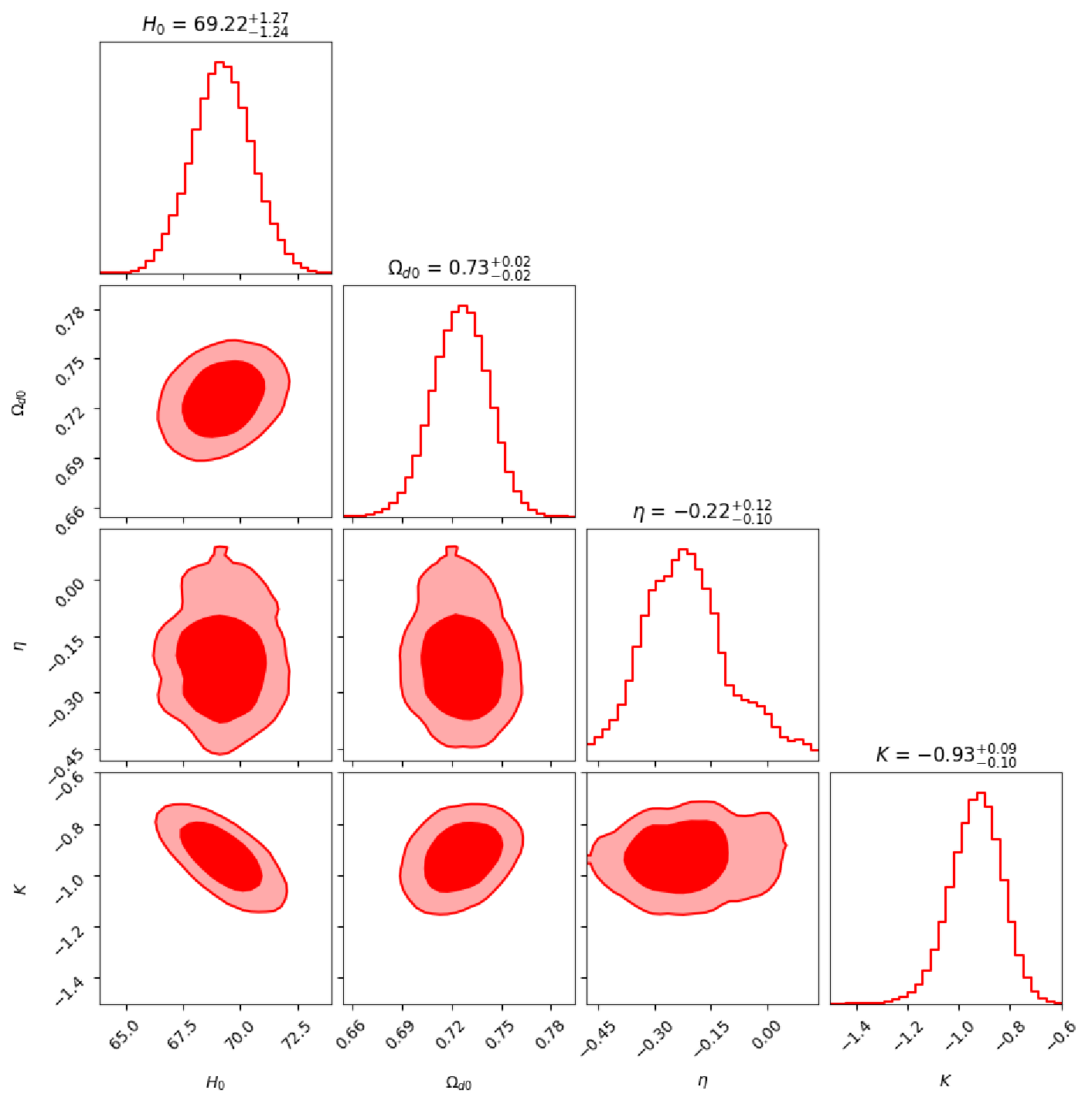}
  \caption{Contour plot for 77 Hubble + 26 BAO data-sets showing 65\% and 95\% confidence levels.}\label{cornerplot}
\end{figure}

\begin{figure}[h!]
  \centering
    \includegraphics[scale=0.3]{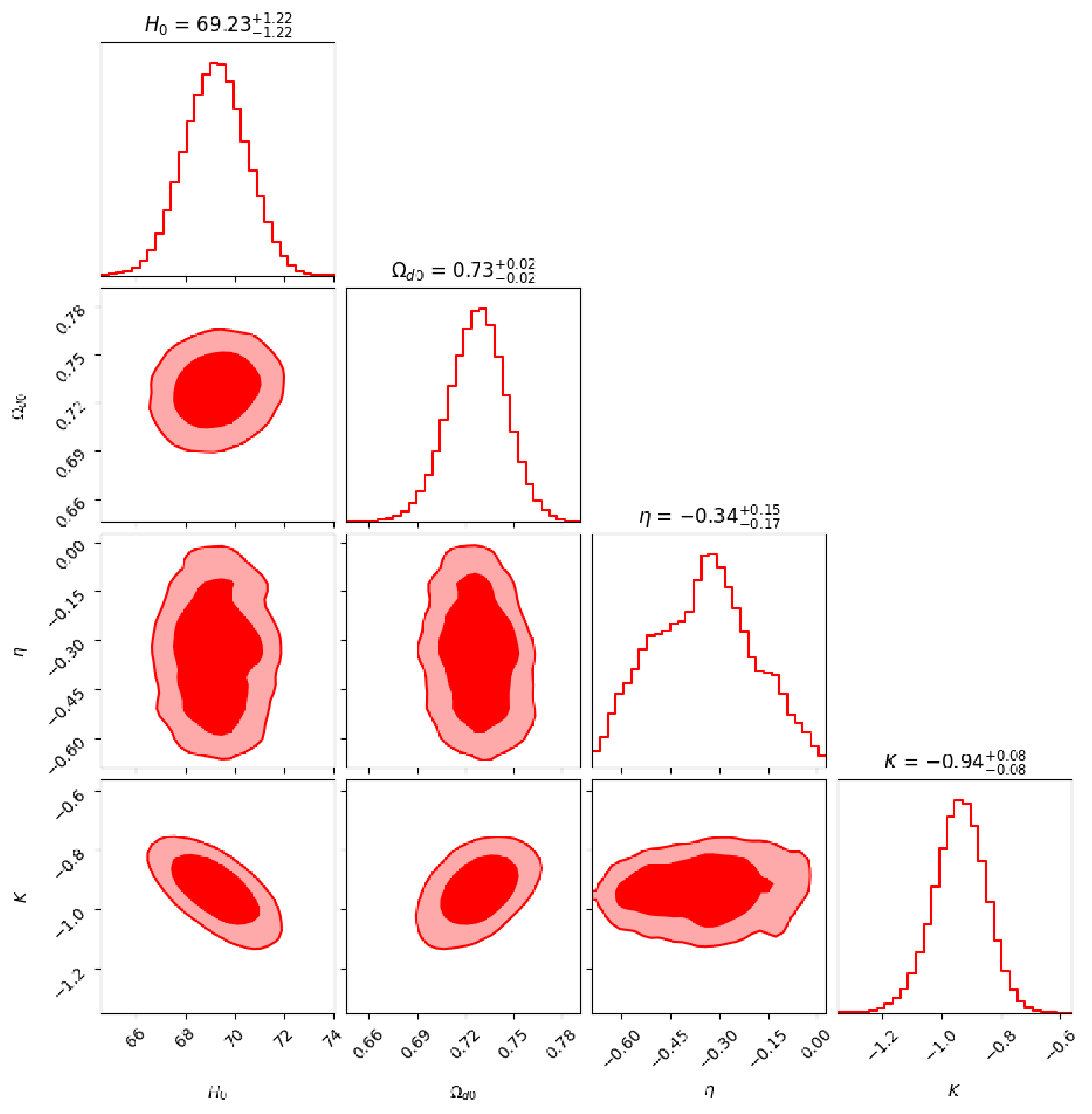}
  \caption{Contour plot for 77 Hubble + 1701 Pantheon$^{+}$ data-sets showing 65\% and 95\% confidence levels.}\label{cornerplot1}
\end{figure}

\begin{figure}[h!]
  \centering
    \includegraphics[scale=0.3]{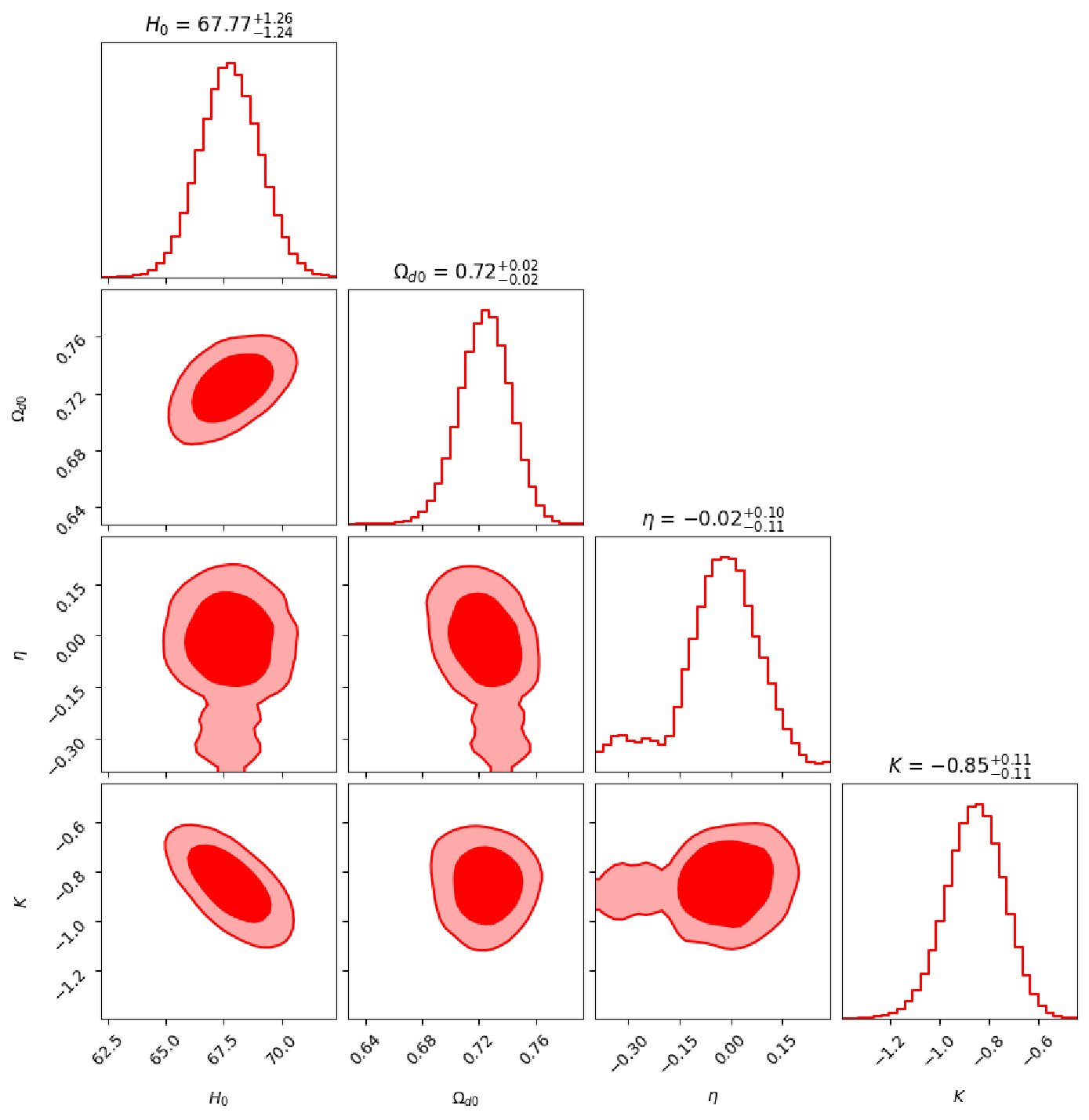}
  \caption{Contour plot for 77 Hubble + 26 BAO + DESI DR2 data-sets showing 65\% and 95\% confidence levels.}\label{cornerplot2}
\end{figure}

\begin{figure}[h!]
  \centering
  \begin{subfigure}[b]{0.45\textwidth}
    \includegraphics[width=\linewidth]{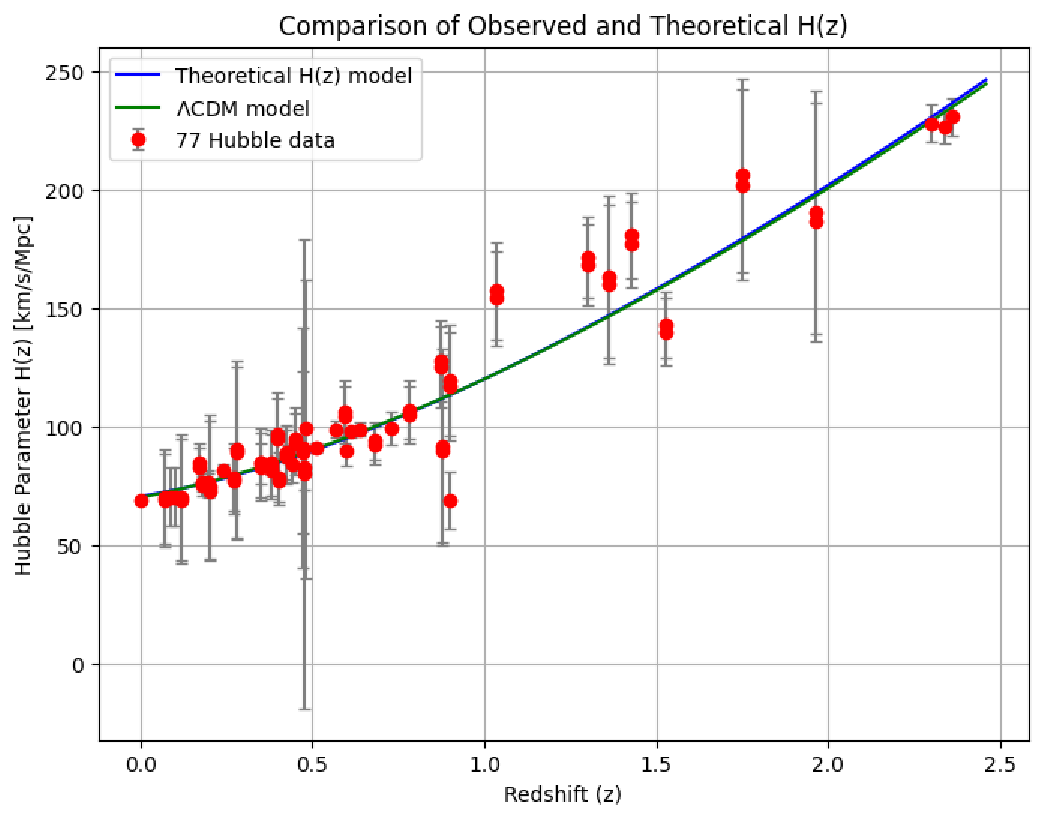}
    \caption{}
    \label{fig:hub}
  \end{subfigure}
  \hfil
  \begin{subfigure}[b]{0.45\textwidth}
    \includegraphics[width=\linewidth]{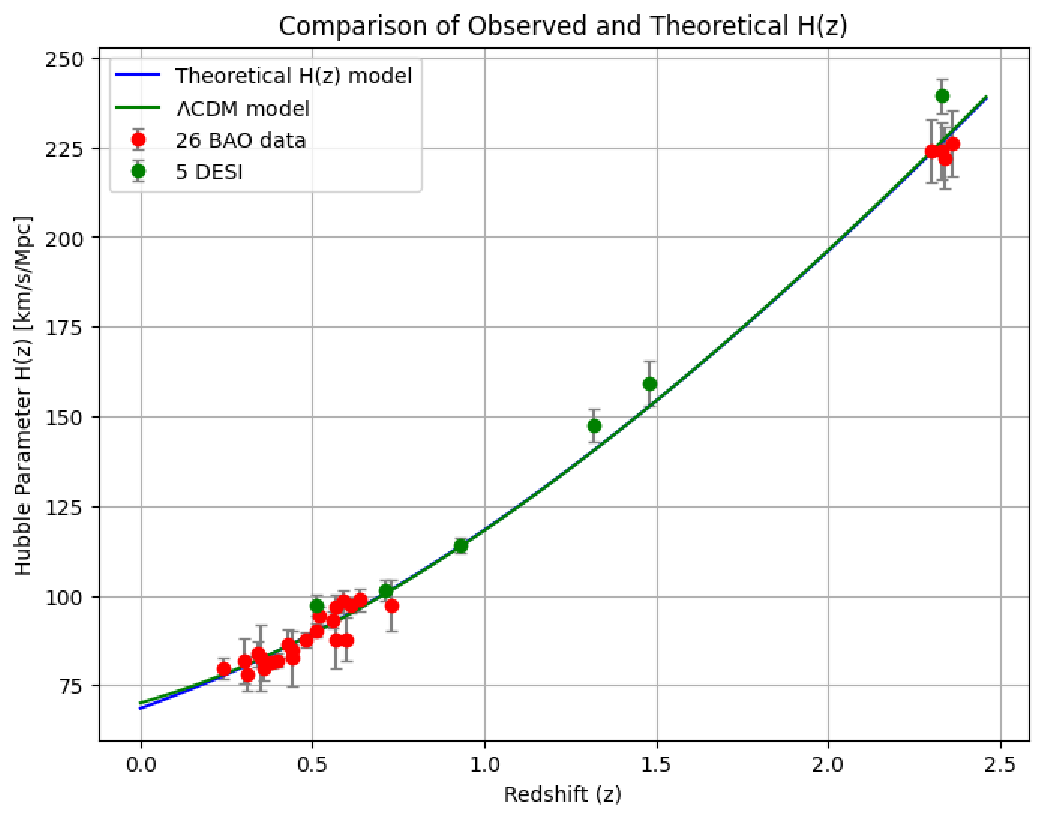}
    \caption{}
    \label{fig:bao}
  \end{subfigure}
  \hfil
  \begin{subfigure}[b]{0.5\textwidth}
    \includegraphics[width=\linewidth]{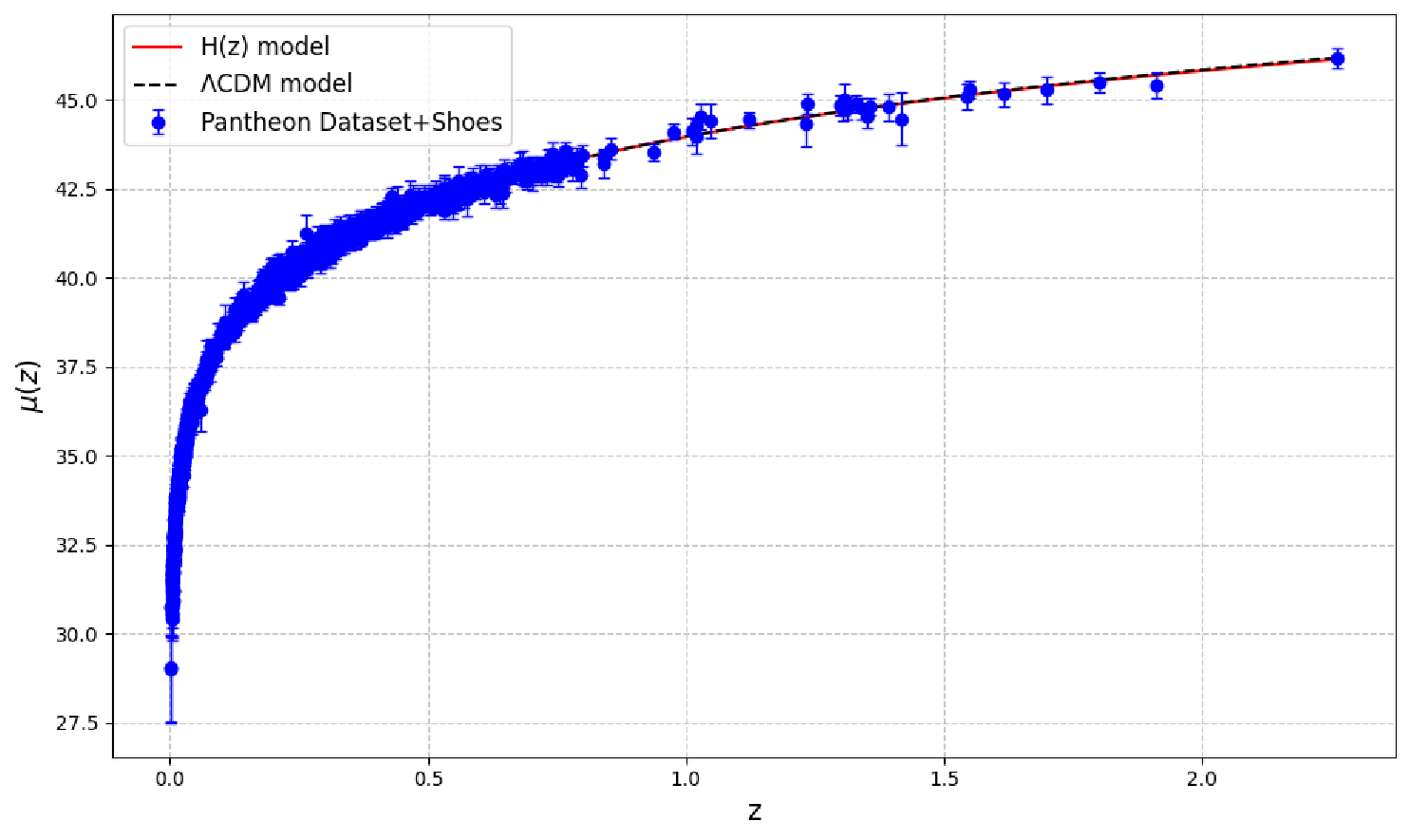}
    \caption{}
    \label{fig:pan}
  \end{subfigure}
  \caption{(a), (b) and (c) show comparison of our model with the $\Lambda$CDM model $H_{0}=69$, $\Omega_{d0}=0.73$, $\eta=-0.34$ and $K=-0.93$}
\end{figure}

Figures~(\ref{cornerplot}), (\ref{cornerplot1}), and (\ref{cornerplot2}) present the 65\% and 95\% confidence level contour plots for the model parameters, obtained using the Hubble77+BAO26, Hubble77+Pantheon$^{+}$, and Hubble77+BAO26+DESI DR2 datasets, respectively. For the Hubble77+BAO26 dataset, the best-fit values of the parameters are $H_0 = 69.22^{+1.27}_{-1.24}$ km\,s$^{-1}$\,Mpc$^{-1}$, $\Omega_{d0} = 0.73^{+0.02}_{-0.02}$, $\eta = -0.22^{+0.10}_{-0.12}$, and $K = -0.93^{+0.09}_{-0.10}$. For the Hubble77+Pantheon$^{+}$ dataset, we obtain $H_0 = 69.23^{+1.27}_{-1.22}$ km\,s$^{-1}$\,Mpc$^{-1}$, $\Omega_{d0} = 0.73^{+0.02}_{-0.02}$, $\eta = -0.34^{+0.15}_{-0.17}$, and $K = -0.94^{+0.08}_{-0.08}$. Finally, for the combined Hubble77+BAO26+DESI DR2 dataset, the best-fit values are $H_0 = 67.77^{+1.26}_{-1.24}$ km\,s$^{-1}$\,Mpc$^{-1}$, $\Omega_{d0} = 0.73^{+0.02}_{-0.02}$, $\eta = -0.02^{+0.10}_{-0.11}$, and $K = -0.85^{+0.11}_{-0.11}$. These results indicate that the contribution of polytropic dark energy to the gravitational action is consistent with the present cosmic epoch. The best-fit values of $H_0$ are in agreement with the Planck 2018 CMB estimate of $H_0 \approx 67.0 \pm 0.5$ km\,s$^{-1}$\,Mpc$^{-1}$ \cite{ref54}, and the present value $\Omega_{d0} \approx 0.73$ is also consistent with current observational constraints. The negative values of the interaction parameter $\eta$ suggest a transfer of energy from dark energy to matter, consistent with previous findings \cite{N roy}. The evolution of the Hubble parameter in our model is compared with that of the $\Lambda$CDM model in Figures~(\ref{fig:hub}) and (\ref{fig:bao}), using the 77 Hubble parameter measurements and the combined 26 BAO + DESI DR2 measurements, respectively. In both cases, the predicted $H(z)$ closely follows the behaviour of $\Lambda$CDM. Furthermore, Figure~(\ref{fig:pan}) presents the comparison of the distance modulus with the Pantheon$^{+}$ dataset comprising 1701 supernovae, demonstrating that our model reproduces the same qualitative trend as the $\Lambda$CDM model.

\begin{table}[h!]
\centering
\begin{tabular}{|c|c|c|c|c|}
\hline
Data-sets  & $H_0$  & $\Omega_{d0}$  & $\eta$ & $K$ \\
\hline
Hubble77+BAO26 & $69.22_{-1.24}^{+1.27}$ & $0.73_{-0.02}^{+0.02}$ & $-0.22_{-0.12}^{+0.10}$ & $-0.93_{-0.10}^{+0.09}$ \\
\hline
Hubble77+pantheon$^+1701$ & $69.23_{-1.22}^{+1.22}$ & $0.73_{-0.02}^{+0.02}$ & $-0.34_{-0.17}^{+0.15}$ & $-0.94_{-0.08}^{+0.08}$ \\
\hline
Hubble77+BAO26+DESI DR2 & $67.77_{-1.24}^{+1.26}$ & $0.72_{-0.02}^{+0.02}$ & $-0.02_{-0.11}^{+0.10}$ & $-0.85_{-0.11}^{+0.11}$ \\
\hline
\end{tabular}
\caption{Values of the constrains in $H(z)$, obtained by using MCMC analysis}
\label{Tparameter}
\end{table}

\section{Analysis of cosmic parameters in the interacting scenario}\label{cosmic parameters}
\hspace{0.5cm}Here, in this section, we do analysis of cosmic parameters, such as energy density, pressure, EoS parameter, deceleration parameter and density parameters, by using the best-fit values obtained from MCMC analysis.

\begin{figure}[h!]
  \centering
  \begin{subfigure}[b]{0.45\textwidth}
    \includegraphics[width=\linewidth]{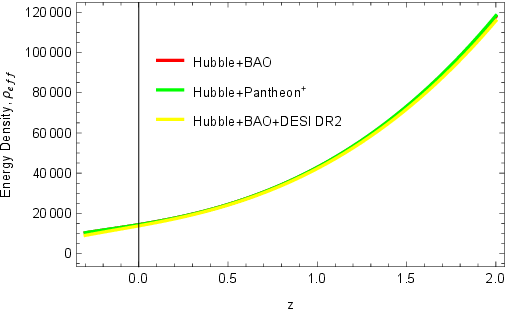}
  \end{subfigure}
  \hfil
  \begin{subfigure}[b]{0.45\textwidth}
    \includegraphics[width=\linewidth]{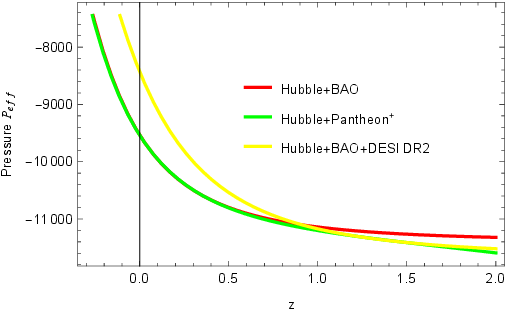}
  \end{subfigure}
  \caption{Evolution of energy density (left) and pressure (right) vs $z$ for all the three combinations of datasets.}\label{ep}
\end{figure}

\textbf{I. Energy density and pressure:} Pressure and energy density are key quantities used to describe the different components of the Universe such as matter, radiation, and dark energy and how they influence the expansion of the Universe via the Friedmann equations. Here, in this section, we study the energy density, $\rho_{eff}$ and pressure, $P_{eff}$ of the Universe. As, we consider matter to be incompressible, the pressure is nothing but the pressure of the polytropic dark energy. For the present context, using equations (\ref{form}), (\ref{rhoM2}) and (\ref{rhoD1}), their expressions are given by

\begin{equation}\label{ed}
   \rho_{eff}(z)  = 3H^{2}_{0}\left[(1-\Omega_{d0})(1+z)^{3}\left[\left(1+K\right)(1+z)^{-3}-K\right]^{\eta}+\Omega_{d0}\left[\left(1+K\right)(1+z)^{-3}-K\right]^{\eta-1}\right]
 \end{equation}
 
 and
 {\footnotesize
 \begin{equation}\label{edp}
 \begin{split}
       &P_{eff}(z) =-3H^{2}_{0}\left[(1-\Omega_{d0})(1+z)^{3}\left[\left(1+K\right)(1+z)^{-3}-K\right]^{\eta}+\Omega_{d0}\left[\left(1+K\right)(1+z)^{-3}-K\right]^{\eta-1}\right]\\
       &+\frac{3 (1 - \Omega_{d0}) (1 + z)^2 \left(-K + \frac{1 + K}{(1 + z)^3}\right)^\eta 
       - \frac{3 \Omega_{d0} (1 + K) (-1 + \eta) \left(-K + \frac{1 + K}{(1 + z)^3}\right)^{-2 + \eta}}{(1 + z)^4}
       -\frac{3 (1 - \Omega_{d0}) (1 + K) \eta \left(-K + \frac{1 + K}{(1 + z)}^3\right)^{-1 + \eta}}{(1 + z)}}
       {(H_0(1+z))^{-1}\sqrt{(1-\Omega_{d0})(1+z)^{3}\left[\left(1+K\right)(1+z)^{-3}-K\right]^{\eta}+\Omega_{d0}\left[\left(1+K\right)(1+z)^{-3}-K\right]^{\eta-1}}}
\end{split}
 \end{equation}
 }
 The evolutions of the energy density and pressure with respect to $z$ are shown in left panel and right panel respectively in Figure (\ref{ep}). It is observed that the energy density remains positive through $z$ and pressure remains negative through $z$ for all the three combinations of the datasets. These negative pressure is responsible for the accelerated expansion of the Universe. These observations are resembled to that of the cosmological constant in the $\Lambda$CDM model.  
 
\textbf{II. EoS parameter:} It is defined as the ratio of total pressure to that of the total energy density. It is represented by $\omega$ and is related to Hubble parameter as $\omega=-1-\frac{\dot{2H}}{3H^2}$. This parameter plays a crucial role in determining how different components of the Universe evolve with time. The value $\omega=-1$ represent $\Lambda$CDM dark energy, $-1<\omega<-\frac{1}{3}$ represent Quintessence, $\omega<-1$ represent phantom dark energy. For the present context, using equation (\ref{effeos}), $\omega_{eff}$ is given by

\begin{equation}\label{EoS1}
\begin{split}
  \omega_{eff}(z)=&\frac{\Omega_{d0} K (-1 + 3 (-1 + \eta + K\eta) z + 
    3 (-1 + \eta + K \eta) z^2 + (-1 + \eta + K \eta) z^3)}{(-1 + 
   K z (3 + 3 z + z^2)) (1 + (-1 + \Omega_{d0}) K z (3 + 3 z + z^2))}\\
   &-\frac{ 
 \eta (-1 + K^2 z (3 + 3 z + z^2) + K (-1 + 3 z + 3 z^2 + z^3))}{(-1 + 
   K z (3 + 3 z + z^2)) (1 + (-1 + \Omega_{d0}) K z (3 + 3 z + z^2))}
   \end{split}
\end{equation}

\begin{figure}[htbp]
  \centering
  \begin{subfigure}[b]{0.45\textwidth}
    \includegraphics[width=\linewidth]{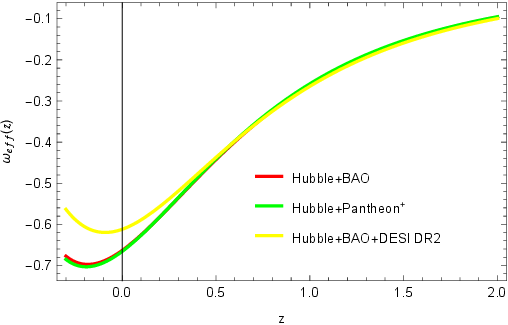}
  \end{subfigure}
  \hfil
  \begin{subfigure}[b]{0.45\textwidth}
    \includegraphics[width=\linewidth]{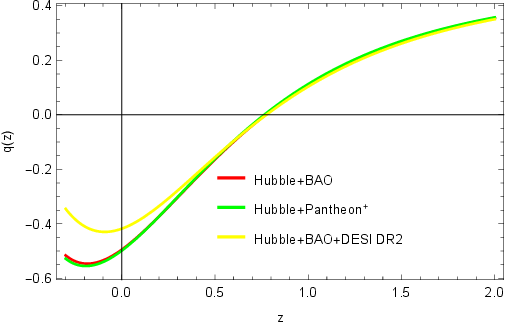}
  \end{subfigure}
  \caption{Evolution of the cosmic parameters EoS (left) and deceleration parameter (right) with respect to $z$ for all the three combinations of datasets.}\label{cp}
\end{figure}

Evolution of $\omega_{eff}$ is shown in left panel of figure (\ref{cp}) . From $z\approx0.8$, the value of $\omega_{eff}$ lie in the interval $(-1,-1/3)$, which correspond to Quintessence era of the Universe. The value of $\omega_{eff}(z)$ at $z=0$ is given in Table \ref{tableA}. These values are found to be $-0.6635$, $-0.6658$ and $-0.612$ for the three combinations of datasets. In all the cases, our Universe is in the phase of Quintessence as the present values lie between $-1$ to $-1/3$. These observation is in agreement with the current conservation, validating the proposed model.

\textbf{III. Deceleration parameter:} It gives the rate at which Universe's expansion is going on. It is denoted by '$q$' and is related to the Hubble parameter as $q=-1-\frac{\dot{H}}{H^2}$. A positive value, a negative value and a zero value indicate that the rate of Universe's expansion is accelerating, decelerating and constant respectively. For the present context, using equation (\ref{decc}), $q$ is given by

\begin{equation}\label{EoS1}
  \begin{split}
     &q(z) = -\frac{(-1 + K z (3 + 3 z + z^2)) (-1 + 3 (1 + K) \eta + K z (3 + 3 z + z^2))+\Omega_{d0} K (-3 + 3 (-4 + 3 (1 + K) \eta) z )}{2 (-1 + K z (3 + 3 z + z^2)) (1 + (-1 + \Omega_{d0}) K z (3 + 3 z + z^2))}\\
     & +\frac{\Omega_{d0} K (
   3 (-4 + 3 \eta + 3 K (1 + \eta)) z^2 + (-4 + 3 \eta + 3 K (6 + \eta)) z^3 + 
   15 K z^4 + 6 K z^5 + K z^6)}{2 (-1 + K z (3 + 3 z + z^2)) (1 + (-1 + \Omega_{d0}) K z (3 + 3 z + z^2))} 
  \end{split}
\end{equation}

The evolution of $q$ is shown in right panel of figure (\ref{cp}). In all the cases, it shows that the values of $q$ starts from a positive values, (due to dominance of matter in the early-time) at higher value of $z$ and becomes negative values, (due to dominance of dark energy in the late-time) at lower values of $z$. The present values of $q$ are given in Table \ref{tableA}. The present value and transition points (point at which the Universe is going from a decelerating phase to accelerating phase) of $q$ are found to be $-0.49525$, $-0.4987$, $-0.418$ and $0.77019$, $0.764184$, $0.77544$ respectively for the three combinations of the datasets. The present values of $q$ are found to be negative, indicating our Universe's expansion is currently accelerating. 
  
\textbf{IV. Density parameters:} They are the dimensionless quantities that represent the ratio of the actual density of the Universe to the critical density required for a flat Universe. The density parameters of matter $\Omega_{m}$ and dark energy $\Omega_{d}$ are key to understand the Universe's composition and evolution. The density parameters of matter and dark energy are given by 

\begin{equation}\label{dMATTER}
  \Omega_{m}(z)=\frac{(1 - K z (3 + 3 z + z^2))(1-\Omega_{d0})}{1 + (-1 + \Omega_{d0}) K z (3 + 3 z + z^2)}
\end{equation}
\begin{equation}\label{dDARK}
  \Omega_{d}(z)=\frac{\Omega_{d0}}{1 + (-1 + \Omega_{d0}) K z (3 + 3 z + z^2)}
\end{equation}

\begin{figure}[h!]
  \centering
  \includegraphics[scale=0.9]{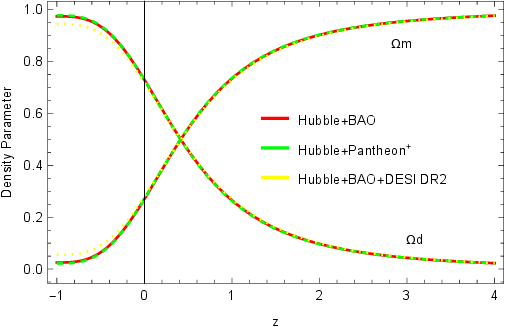}
  \caption{Evolution of density parameters with respect to $z$ for all the three combinations of datasets.}\label{dps}
\end{figure}

From Figure (\ref{dps}), it is seen that in the early phase of the Universe matter is dominated over dark energy. As $z$ approaches $-1$, the density parameter of matter decreases over time and the density parameter of dark energy increases over time. The present values of $\Omega_{m}$ and $\Omega_{d}$ are given in Table \ref{tableA}. These values are found to be 0.27, 0.27, 0.28 and 0.73, 0.73, 0.72 respectively for the $\Omega_{m0}$ and $\Omega_{d0}$ respectively for the three combinations of datasets. In all the cases, the values of $\Omega_{d0}$ are found to be larger than that of the $\Omega_{m0}$, indicating presently, our Universe is dominated by dark energy, which corresponds to the accelerated expansion of the Universe. These observations are closely agreement with the current observational data, indicating our model has the ability to mark out the full evolution of the Universe and its composition.

\section{State-finder diagnostic}\label{sf}

 \hspace{0.5cm}The state-finder diagnostic is a tool introduced to characterise and differentiate between various dark energy models that explain the observed accelerated expansion of the Universe. While the Hubble parameter $H(z)$ and deceleration parameter $q(z)$ describe the first and second derivatives of the scale factor 
$a(t)$, the state-finder involves higher-order derivatives, providing a more detailed picture of cosmic dynamics. Defined through the parameters $\{r,s\}$, where 
$r=\frac{\dddot{a}}{3H^3}$ and $s=\frac{r-1}{3(q-1/2)}$, this pair allows for a more refined classification of cosmological models. In this formalism, the standard $\Lambda$CDM model corresponds to the fixed point $\{r=1,s=0\}$, serving as a reference for comparison. Deviations from this point reveal the dynamical nature of alternative dark energy or modified gravity theories. For instance, quintessence models typically yield $r<1$ and $s>0$, while phantom models often have 
$r>1$ and $s<0$. Thus, the trajectory of a model in the $r-s$ plane reveals both its present behaviour and its evolution, offering a geometric and quantitative tool for assessing its viability against cosmological observations.

\begin{figure}[h!]
  \centering
  \includegraphics[scale=1]{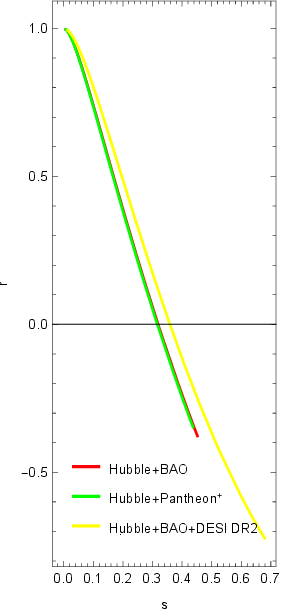}
  \caption{Plot of the state-finder parameters for all the three combinations of datasets.}\label{sfp}
\end{figure}

The evolution of state-finder parameters are given in Figure \ref{sfp}. In all the cases, they approach to the $\Lambda$CDM model at the late-time. The present values of $\{r,s\}$ are given in Table \ref{tableA}. These values are found to be \{0.484116,0.172782\}, \{0.510846,0.163263\} and \{0.1738,0.3\} respectively for the three combinations of datasets. It is found that the present values of $r(z)$ are less than 1 and that of the $s(z)$ is greater than 0, indicating our Universe is in the phase of Quintessence, indicating slower acceleration growth than in $\Lambda$CDM.

\begin{table}[h!]
\centering
\begin{tabular}{|c|c|c|c|c|c|c|}
\hline
Data-sets & $\omega_{eff}$ & $q$  & $\Omega_{m0}$ & $\Omega_{d}$ & $\{r,s\}$ & Age \\
\hline
Hubble77+BAO26 & -0.6635 & -0.49525  & 0.27 & 0.73  & \{0.484116,0.172782\} & 14.16 \\
\hline
Hubble77+Pantheon$^+1701$ & -0.6658 & -0.4987 & 0.27  & 0.73 & \{0.510846,0.163263\}&14.15\\
\hline
Hubble77+BAO26+DESI DR2 & -0.6120 & -0.4180  & 0.28 & 0.72 & \{0.1738,0.3\}&14.10 \\
\hline
\end{tabular}
\caption{Present values of the $\omega_{eff}$, $q$, $\Omega_{m0}$, $\Omega_{d}$, $\{r,s\}$ and age of the Universe for all the combinations of the datasets.}
\label{tableA}
\end{table}

\section{Age of the Universe:}\label{age} 

\begin{figure}[htbp] 
   \centering
   \includegraphics[scale=0.9]{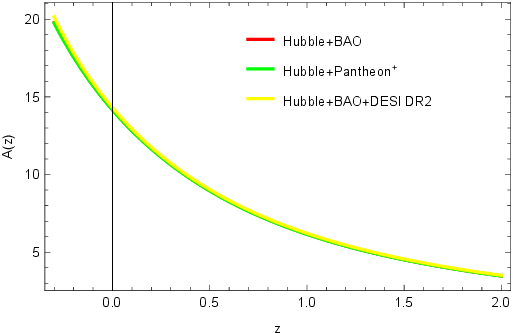} 
   \caption{Plot of the age of the Universe vs $z$ for all the three combinations of datasets.}
   \label{A}
\end{figure}

\hspace{0.5cm}The validity of the model can be tested by using the age of the Universe. The determination of the cosmic age provides a fundamental test of cosmological models, as it encapsulates the integrated expansion history of the Universe from the Big-Bang to the present epoch. The age of the Universe, $A(z)$ in term of redshift is defined as
\begin{equation}\label{ageEx}
  A(z)=\int_{z}^{\infty}\frac{dx}{(1+x)H(x)}
\end{equation}
In figure (\ref{A}), the plot of the $A(z)$ with respect to $z$ is presented. The present age of the Universe is given in Table \ref{tableA}. The present values are found to be 14.16, 14.15 and 14.10 respectively for the three combinations of the datasets. These values are closed with current age 13.86 Gyr obtained from the $\Lambda$CDM model. This closeness of the values of our model to that of the $\Lambda$CDM model, suggests that our model is observational validated.

\section{Behaviour of the interacting term}\label{behaviour}
\begin{figure}[htbp]
  \centering
  \includegraphics[scale=0.9]{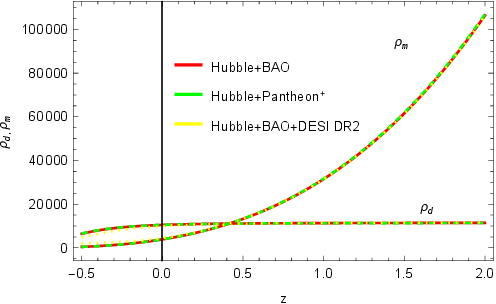}
  \caption{Evolution of energy density of matter and dark energy vs $z$ for all the three combinations of datasets.}\label{edc}
\end{figure}

\hspace{0.5cm}In the interacting term, $Q=3\eta H \rho_{d}$, $\eta$ plays a crucial role in examining the direction of transferred of energy, whether it is from dark energy to matter or matter to dark energy. From the MCMC analysis, the values of $\eta$ is found to be negative for all the three combinations of data-sets, indicating there is transferred of energy from dark energy to matter. But one expect that transferring of energy takes from matter to dark energy \cite{G. Olivares}\cite{Zimdahi}. Present case of transferring of energy is also possible as seen in \cite{N roy}. To check the validity of the reverse flow of energy, we need to look for  the expression of energy density of both matter and dark energy given in equations (\ref{rhoM2}) and (\ref{rhoD1}) respectively. The evolution of these energy density is shown in figure (\ref{edc}). It is observed that in spite of the energy flow from the dark energy to dark matter, the present value of $\rho_d$ is larger than that of the $\rho_m$ and also in the future. The reason might be due to weak strength of the interaction and the evolution of the Hubble parameter present in the interacting term.  

\newpage
\section{Discussion}\label{dis}
\hspace{0.5cm}In this current study, we explored polytropic dark energy, $p_d=\alpha \rho_{d}^{1+\frac{1}{\beta}}$, in an interacting scenario with a simple form of the interacting term $Q=\eta H\rho_{d}$. In this condition, the Hubble parameter is obtained as  $H(z)=H_{0}\left[(1-\Omega_{d0})(1+z)^{3}\left[\left(1+K\right)(1+z)^{-3}-K\right]^{\eta}+\Omega_{d0}\left[\left(1+K\right)(1+z)^{-3}-K\right]^{\eta-1}\right]^{1/2}$. The optimal values of the parameters in our model are constrained by using MCMC analysis using Hubble77+BAO26, Hubble77+Pantheon$^{+}$ and Hubble77+BAO26+DESI DR2. With respect to these three combinations of data-sets, we obtained the values of $H_{0}$ as (69.22, 69.23, 67.77); the values of $\Omega_{d0}$ as (0.73, 0.73, 0.72); the values of $\eta$ as ($-0.22$, $-0.34$, $-0.02$) and the values of $K$ as ($-0.93$, $-0.94$, $-0.85$) respectively. We compare the proposed model with the standard $\Lambda$CDM scenario by graphically examining the evolution of various cosmological parameters as functions of the redshift $z$. Across the entire $z$ range, the energy density remains positive, while the pressure remains negative, thereby sustaining cosmic acceleration. For the three considered combinations of observational datasets, the present-day values of the effective equation-of-state parameter $\omega_{\mathrm{eff}}$ are obtained as $-0.6635$, $-0.6658$, and $-0.612$, respectively, indicating that the Universe resides in a Quintessence phase. Similarly, the corresponding deceleration parameter values, $q = -0.49525$, $-0.4987$, and $-0.418$, confirm the ongoing accelerated expansion. The evolution of the density parameters reveals a monotonic decrease in the matter density parameter over time, while the dark energy density parameter increases, signifying present-day dark energy dominance. The state-finder diagnostics yield $r(0) < 1$ and $s(0) > 0$ for all cases, again consistent with Quintessence behaviour. The estimated present ages of the Universe, $14.16$, $14.15$, and $14.10$ Gyr for the three dataset combinations, are in close agreement with $\Lambda$CDM predictions. Additionally, we investigate the interaction term $Q = 3\eta H \rho_d$ and find $\eta < 0$ in all cases, implying a transfer of energy from dark energy to matter. Overall, these results suggest that the model provides a viable alternative to $\Lambda$CDM, capable of explaining late-time cosmic acceleration while remaining observationally consistent.

 \end{document}